# Three photon absorption in ZnO and ZnS crystals

**Jun He, Yingli Qu, Heping Li, Jun Mi, and Wei Ji[*]**

*Department of Physics, National University of Singapore, 2 Science Drive 3, Singapore 117542, Republic of Singapore*
*phyjiwei@nus.edu.sg*

**Abstract:** We report a systematic investigation of both three-photon absorption (3PA) spectra and wavelength dispersions of Kerr-type nonlinear refraction in wide-gap semiconductors. The Z-scan measurements are recorded for both ZnO and ZnS with femtosecond laser pulses. While the wavelength dispersions of the Kerr nonlinearity are in agreement with a two-band model, the wavelength dependences of the 3PA are found to be given by $(3E_{\text{photon}}/E_g - 1)^{5/2}(3E_{\text{photon}}/E_g)^{-9}$. We also evaluate higher-order nonlinear optical effects including the fifth-order instantaneous nonlinear refraction associated with virtual three-photon transitions, and effectively seventh-order nonlinear processes induced by three-photon-excited free charge carriers. These higher-order nonlinear effects are insignificant with laser excitation irradiances up to 40 GW/cm$^2$. Both pump-probe measurements and three-photon figures of merits demonstrate that ZnO and ZnS should be a promising candidate for optical switching applications at telecommunication wavelengths.

## 1. Introduction

As a wide and direct band gap (~ 3.3 eV) semiconductor, zinc oxide (ZnO) is one of the materials that have tremendous potentials for blue-ultraviolet light emitters and detectors, transparent high-power electronics, and piezoelectric transducers. Its low threshold for optical pumping and large exciton binding energy (~ 60 meV) allow lasing action in ZnO to be reached at extremely low pumping power at room temperature. Because of these potential applications, extensive research efforts have been made on the basic physical properties of ZnO. A comprehensive review on the synthesis, mechanical, chemical, thermal and optical properties of ZnO may be found in Ref. [1]. Similarly, wide-band-gap zinc sulfide (ZnS) is also expected to play an equally important role in optoelectronic devices. ZnS-based thin films for both field emission displays and photovoltaic applications have been reported [2,3]. And ZnS-based photonic crystals are also proposed [4].

There have been reports on the nonlinear optical properties of ZnO and ZnS, including two-photon absorption (2PA), Kerr-type refractive nonlinearity, and nonlinear processes induced by 2PA-excited charge carriers [5-7]. Recently, ultrafast carrier dynamics in nano-structured ZnO [8,9], and huge nonlinear refraction and absorption in ZnO thin films have been observed [10]. Ultrafast all-optical switching devices are the key component for next generation broadband optical networks. The implementation of such devices requires materials with low linear and nonlinear losses, high Kerr-type refractive nonlinearities and response times of a few picoseconds or less [11]. To search for materials which meet the above requirements, it has been suggested to target at semiconductors with their band gap at least twice the photon energy used ($E_g > 2 E_{photon}$) avoiding optical absorption due to one- or two-photon transitions [12-14]. In this regard, wide-gap semiconductors are a suitable candidate; and ZnSe has been recently reported to be one of the promising candidates [15]. When a semiconductor device is operated at photon energies below half the band gap, three-photon absorption (3PA) appears to become a limiting factor for all-optical switching. In addition, there has been an increasing interest in using 3PA of wide-gap semiconductor quantum dots for biological imaging [16]. However, there is a lack of direct measurements on the 3PA spectra of wide-gap semiconductors, and reports in the literature are sporadic [17,18]. Furthermore, the old measurements were performed using laser pulses of nanosecond or picosecond duration. There was a possibility that other nonlinear processes like excited charge carriers may contribute to the observed 3PA process. Such a possibility could be minimized if femtosecond laser pulses are utilized.

In this paper we report a systematical investigation into both 3PA and Kerr-type refractive nonlinearity ($n_2$) in wide-gap semiconductors, ZnO and ZnS. In particular, the 3PA spectra and $n_2$ wavelength dispersions are probed in detail with both femtosecond Z-scan and time-resolved pump-probe techniques. We have found that the 3PA spectra of wide-gap semiconductors fit well to a model developed by Brandi and de Araujo [19], contradicting the results for narrow-band-gap semiconductors [14]. This paper is organized as follows. First, the experimental details are described in Section 2. This is followed by a theoretical analysis for Z-scans on the nonlinear medium with 3PA or 2PA. In Section 4, comparison between our experimental data and theories on the wavelength dispersion of both 3PA and $n_2$-nonlinearity is presented. By using both theoretical simulation and experimentally measured pump-probe data, in Section 4, we also give out the justification why higher-order nonlinear mechanisms like the fifth-order nonlinear refraction ($n_4$) and 3PA-excited charge carrier effects can be ignored under our experimental condition. Finally, the three-photon figures of merit are assessed for ZnO and ZnS.

## 2. Experimental

The 3PA properties of two single crystals (ZnO: hexagonal structure, <0001> orientation, 10×10×1.0 mm in size, Atramet Inc; and ZnS: cubic structure, <111> orientation, 10×10×0.5 mm in size, Semiconductor Wafer Inc.) were determined by Z-scans at room temperature with laser pulses of wavelengths ranging from 720 to 950 nm. The laser pulses were generated by a mode-locked Ti:Sapphire laser (Chameleon, Coherent) operating at a repetition rate of 90 MHz with a pulse duration of 90 ~ 150 fs, depending on the wavelength. The laser pulses were focused with a minimum beam waist of ~ 11 $\mu$m, depending on the wavelength. The crystal was moved along the laser propagation direction (Z-axis). In an open-aperture (OA) configuration, the total transmitted pulse energy was measured in the far field as a function of the crystal position; while only central part of the transmitted pulse energy was recorded for a closed-aperture (CA) Z-scan. The 3PA coefficient and Kerr-type refractive nonlinearity can be unambiguously determined by OA and CA Z-scans, respectively, as discussed in Sections 3 and 4.

We also conducted Z-scans with 120-fs, 780-nm laser pulses at a lower repetition rate (1 kHz) in order to eliminate the possibility of thermal contribution to the measured 3PA coefficient or Kerr-type refractive nonlinearity. These high-energy laser pulses were delivered by a Ti:Sapphire regenerative amplifier (Titan, Quantronix), which was capable of producing laser pulse energies of up to 1 mJ. With a set of neutral density filters, the pulse energy was attenuated so that maximum laser irradiance within the crystal was kept below ~ 40 GW/cm$^2$ (fluence: ~ 5 mJ/cm$^2$). It should be pointed out that other higher-order nonlinear processes, such as effective seventh-order nonlinear effects induced by 3PA-excited charger carriers, is expected to occur in the higher irradiance regime (greater than 100 GW/cm$^2$). In addition, extremely intense laser pulses may induce a nonlinear phase variation ($\Delta\phi$) so large that the thin-sample approximation for standard Z-scan analysis becomes invalid any more [20, 21]. To avoid these complex problems, therefore, all the data presented in this paper were acquired with laser irradiances below 40 GW/cm$^2$. It should be emphasized that no laser-induced damage was observed within this range of laser irradiance.

The anisotropy of the 3PA coefficients or the $n_2$ values was also scrutinized with Z-scans for the two crystals. No measurable polarization dependence was observed within our experimental error. To further eliminate long-term accumulative nonlinear processes, we performed femtosecond time-resolved transient absorption and optical Kerr effect (OKE) measurements. The transient absorption measurements were carried out in degenerate pump-probe experiments in a cross-polarized configuration with 120-fs laser pulses (780 nm, 1 kHz repetition rate). The detailed description of the set-up is referred to elsewhere [22]. Similarly, the OKE measurements were obtained from the same experimental set-up except that the states of polarization were altered for the pump and probe pulses.

## 3. Z-scan theories on multiphoton absorbers

Under the condition that the photon energy of laser pulses is less than the band-gap energy ($E_{photon} < E_g$), the multi-photon absorption (MPA) coefficients may be measured by conducting Z-scans. Figure 1(a) illustrates the OA Z-scan curves for the ZnO crystal at two excitation irradiances ($I_{00}$), where $I_{00}$ is denoted as the peak, on-axis irradiance at the focal point within the sample. $I_{00}$ is related to the incident irradiance by taking Fresnel's surface reflection into consideration. By using both thin sample approximation and slowly varying envelope approximation (SVEA), the wave equation can be separated into two equations: one for the nonlinear phase and the other for the irradiance:

$$\frac{d\Delta\phi}{dz'} = k\sum_{m=2} n_{2m-2} I^{m-1} + k\sigma_r N_{e-h} \quad (1)$$

$$\frac{dI}{dz'} = -(\alpha_0 + \sum_{m=2} \alpha_m I^{m-1} + \sigma_a N_{e-h})I \quad (2)$$

where $k$ is the magnitude of the wave vector in free space; $n_{2m-2}$ is the nonlinear index of refraction ($m = 2$ for the third-order nonlinearity, $m = 3$ for the fifth-order nonlinearity, and so on); $\sigma_r$ and $\sigma_a$ are the refractive and absorptive cross sections of photo-excited charge carriers (or photo-excited electron-hole pairs), respectively; $N_{e-h}$ is the number density of the photo-excited charge carriers, $\alpha_0$ is the linear absorption coefficient, $\alpha_m$ is the MPA coefficient ($m = 2$ for 2PA; $m = 3$ for 3PA, and so on); and $I$ is the irradiance within the sample. If we keep the 2PA term and ignore all other terms on the right side of Eq. (2), we can analytically solve Eq. (2) for OA Z-scans on two-photon absorbers. Similarly, by keeping the 3PA term and ignoring the other terms, we can have an analytical expression for OA Z-scans on three-photon absorbers. By assuming a spatially and temporally Gaussian profile for incoming laser pulses, the normalized energy transmittance, $T_{OA}(z)$, for 2PA and 3PA can be derived as Eq. (3) and Eq. (4), respectively [23],

$$T_{OA}(z) = \frac{1}{\pi^{1/2} q_0} \int_{-\infty}^{\infty} \ln[1 + q_0 \exp(-x^2)] dx \quad (3)$$

$$T_{OA}(z) = \frac{1}{\pi^{1/2} p_0} \int_{-\infty}^{\infty} \ln\left\{[1 + p_0^2 \exp(-2x^2)]^{1/2} + p_0 \exp(-x^2)\right\} dx \qquad (4)$$

where $q_0 = \alpha_2 I_0 L_{eff}$; $p_0 = (2\alpha_3 I_0^2 L'_{eff})^{1/2}$; $I_0 = I_{00}/(1+z^2/z_0^2)$ is the excitation intensity at position z; $z_0 = \pi\omega_0^2/\lambda$ is the Rayleigh range; $\omega_0$ is the minimum beam waist at the focal point ($z = 0$); $\lambda$ is the laser free-space wavelength; $L_{eff} = [1 - \exp(-\alpha_0 L)]/\alpha_0$ and $L'_{eff} = [1 - \exp(-2\alpha_0 L)]/2\alpha_0$ are the effective sample lengths for 2PA and 3PA processes, respectively; and $L$ is the sample length. The 2PA (or 3PA) coefficients can be extracted by the best fit between the above equation and the OA Z-scan curve. For the OA Z-scans in Fig. 1(a), the best fits result in a 3PA coefficient of 0.016 cm$^3$/GW$^2$ for the ZnO crystal.

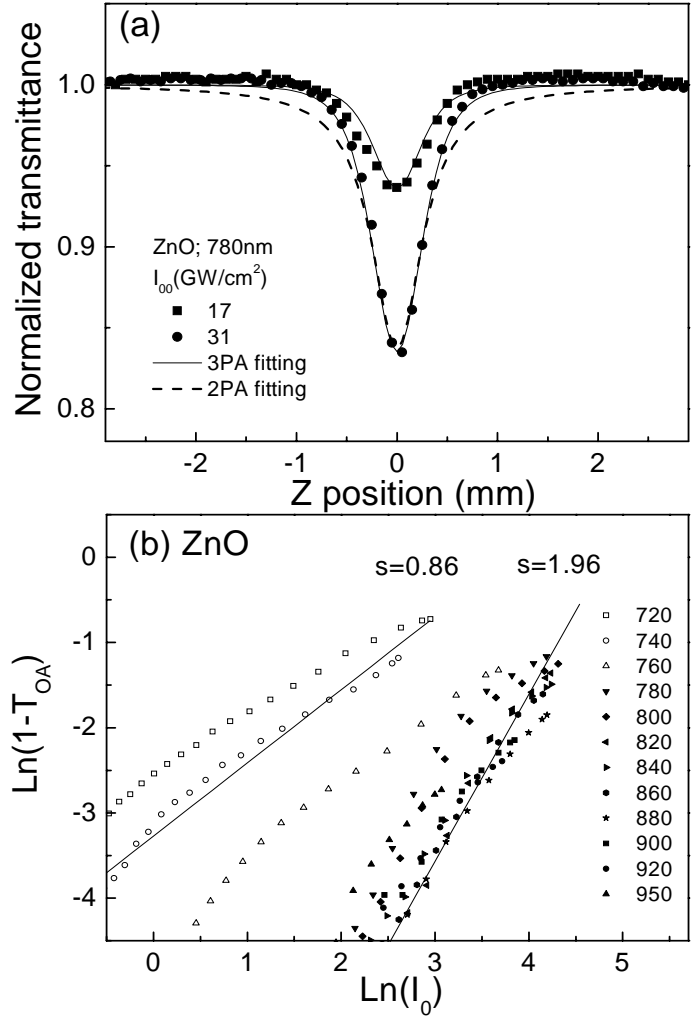

Fig. 1. (a) OA Z-scans measured with different excitation irradiances at a wavelength of 780 nm and a pulse repetition rate of 90 MHz. The solid and dashed lines are the fitting curves by employing the Z-scan theory, described in the text, on 3PA and 2PA respectively. (b) Plots of Ln(1–$T_{OA}$) vs. Ln($I_0$) at different wavelengths, the solid lines are the examples of the linear fit at 740 nm with a slope of s = 0.86 and at 840 nm with a slope of s = 1.96.

It has been pointed out that 3PA Z scans are considerably different from 2PA Z-scans [24]. Our results in Fig. 1(a) are also indicative of large discrepancy between the experiment data and Eq. (3) which is mathematically described only for 2PA. If $q_0 < 1$ or $p_0 < 1$, Eq. (3) or (4) can be expanded in a Taylor series as

$$T_{OA} = \sum_{m=0}^{\infty} (-1)^m \frac{q_0^m}{(m+1)^{3/2}} \tag{5}$$

$$T_{OA} = \sum_{m=1}^{\infty} (-1)^{m-1} \frac{p_0^{2m-2}}{(2m-1)!(2m-1)^{1/2}} \tag{6}$$

Furthermore, if the higher order terms are ignored, we obtain:

$$T_{OA} = 1 - \alpha_2 I_0 L_{eff} / 2^{3/2} \tag{7}$$

$$T_{OA} = 1 - \alpha_3 I_0^2 L'_{eff} / 3^{3/2} \tag{8}$$

These two expressions enable ones to identify 2PA or 3PA processes. By the use of linear fit to the plot of Ln(1–$T_{OA}$) vs. Ln($I_0$) at different wavelengths, as shown in Fig. 1(b), we can distinguish between 2PA and 3PA by the slope ($s$). If the slope is 2 (or 1), it is indicative of 3PA (or 2PA). For the ZnO crystal ($E_g$ = 3.3 eV), as shown in Fig. 1(b), the 2PA process dominates in the wavelength range from 720 to 770 nm ($s$ = 0.84 ~ 0.98), while the 3PA occurs at longer wavelengths ($s$ = 1.78 ~ 1.98). The interband 2PA in the ZnS crystal was unobserved because the photon energies (1.3 ~ 1.7 eV) used in our experiments were less than half of its band-gap energy ($E_g$ = 3.7 eV)

### 4. Results and discussion

*4.1 Three-photon absorption*

In Fig. 2, we plot the extracted 3PA coefficient as a function of the ratio of the photon energy to the band-gap energy ($E_{photon}/E_g$). In Table 1, the 3PA coefficients are also summarized for ZnO and ZnS. For 3PA theories on semiconductors, there have been two approaches under the two-band approximation: the first one, developed by Wherrett [25], uses the first-order perturbation theory with maximum available interband excitations and de-excitations; and the second one includes a single, virtual interband transition and two self-transitions, derived by Brandi and de Araujo [19]. The Wherrett's theory leads to a wavelength scaling of $(3E_{photon}/E_g–1)^{1/2}(3E_{photon}/E_g)^{-9}$, while the second approach results in a different dispersion of $(3E_{photon}/E_g–1)^{5/2}(3E_{photon}/E_g)^{-9}$. The expressions of the two models are given, respectively, as follows:

$$\alpha_3 = \frac{3^{10} 2^{1/2}}{8} \pi^2 \left(\frac{e^2}{\hbar c}\right)^3 \frac{\hbar^2 P^3}{n_0^3 E_g^7} \frac{(3E_{photon}/E_g - 1)^{1/2}}{(3E_{photon}/E_g)^9} \tag{9}$$

$$\alpha_3 = \frac{2^{9/2} 3^{10} \pi^2}{5} \left(\frac{e^2}{\hbar c}\right)^3 \hbar^5 S_3 \frac{(3E_{photon}/E_g - 1)^{5/2}}{(3E_{photon}/E_g)^9} \tag{10}$$

where $S_3 \equiv (p_{vc}^2/m^2)(m^*)^{7/2}/(m_1^4 n_0^3 E_g^{13/2}) \approx (3/4)(m^*)^{5/2}/(m_1^4 n_0^3 E_g^{11/2})$, $m_1$ is the effective mass of the conduction band and $m^*$ is the reduced effective mass which is defined as

$(m^*)^{-1} = (m_1)^{-1} + (|m_0|)^{-1}$. $P$ is the Kane parameter, $n_0$ is the refractive index of the material, and $e^2/\hbar c$ the fine-structure constant. We plot the two expressions in Fig. 2, illustrating that the measured 3PA spectra are close to the theory predicted by Brandi and de Araujo [19]. A 3PA coefficient measured by Catalano et al. is also added for comparison [17]. It is interesting to note that for AlGaAs semiconductor, the Wherrett's theory is experimentally verified to be more accurate [14].

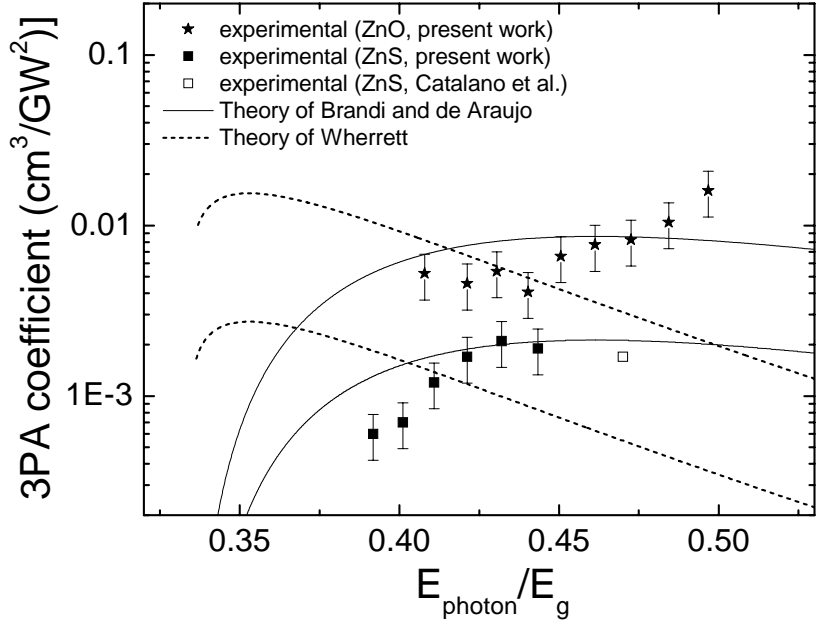

Fig. 2. 3PA coefficients plotted as a function of $E_{photon}/E_g$, where $E_{photon}$ is the photon energy and $E_g$ is the band-gap energy. The solid stars and squares are the experimental data, while the solid and dashed lines are calculated from the theory of Brandi and de Araujo [19] and the theory of Wherrett [25], respectively. For comparison, the experimental result (the empty square) of Catalano et al. is also displayed [17].

We compared the Z-scans measured with 1-kHz pulse repetition rate to the ones obtained with 90-MHz pulse repetition rate. No measurable difference between these two results was found within the experimental error, which ensures that thermal effects were negligible under our experimental condition. Figure 3(a) illustrates typical OA Z-scans at 780 nm at different excitation irradiances by the use of 1-kHz repetition rate laser pulses. The extracted 3PA coefficients are nearly independent of the irradiance, as shown in the inset of Fig. 3(a), confirming the dominance of the fifth-order nonlinear absorption process.

*4.2 Kerr nonlinearity*

Figure 3(b) shows typical CA Z-scans at various excitations. To determine the Kerr-type refractive nonlinearity, we divide the CA Z-scan by the OA Z-scan measured at the same excitation irradiance. Figure 4(a) shows a typical example for such a division. By keep the first term of $n_2 I$ on the right side of Eq. (1), an analytical solution can be obtained. For the detailed derivation, the reader is referred to Ref. [20]. By the best fit with the Z-scan theoretical solution, the values of $n_2$ may be extracted from the Z-scan data. Table 1 summarizes all the extracted $n_2$ values for ZnO and ZnS in the wavelength range of 720-950 nm. The measured values of $n_2$ are compared to the theoretical values obtained from a two parabolic band model based on the Kramers-Kronig (KK) relation [26]. Within an order of magnitude, the $n_2$ data are in agreement with the theory for both ZnO and ZnS crystals, as shown in Fig. 4(b). The factor of 2-3 discrepancy in ZnO crystal can be partially attributed to using a single valence band (light hole) and ignoring the transition originating from heavy-hole valence band. Moreover, as described in Ref. [27], inclusion of electron-hole Coulomb interaction will further improve the agreement between theory and experiment. For comparison, the experimental data of Adair et al. [5], Zhang et al. [6], and Krauss et al.[7], are also included in Fig. 4(b).

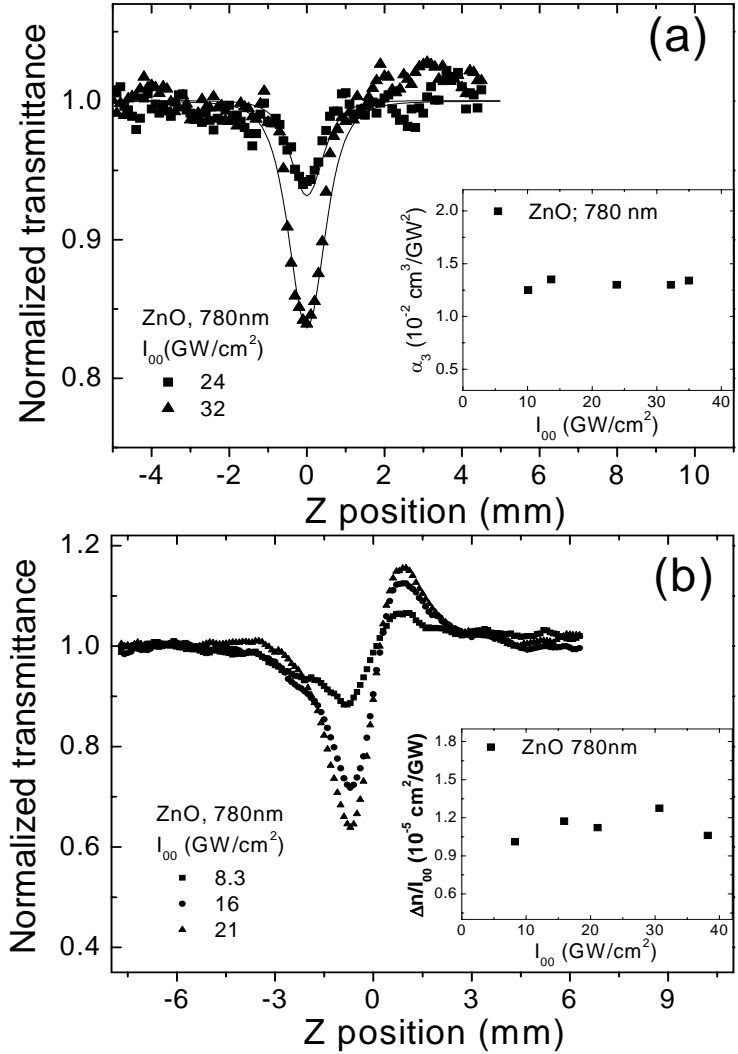

Fig. 3. (a) OA Z-scans and (b) CA Z-scans divided by OA Z-scans measured with 1-kHz repetition rate laser pulses at various excitation irradiances. The inset in (a) and (b) shows the irradiance dependence of the 3PA coefficient and the irradiance dependence of the $n_2$ value, respectively.

The fifth-order refractive nonlinearity ($n_4$) is also evaluated both experimentally and theoretically. In the experiment, we find that the measured $n_2$ value is independent of the irradiance, as shown in the inset of Fig. 3(b), which rules out the occurrence of higher-order refractive nonlinearities. Theoretically, a contribution of $n_4 I^2$ to the nonlinear refraction comes from virtual three-photon transitions. In a similar manner to the attainment of $n_2$ from the $\alpha_2$ spectrum [26], the $n_4$ values may be calculated by using a nonlinear KK transformation of the 3PA spectrum, $\alpha_3$:

$$n_4(\omega_1;\omega_2,\omega_3) = \frac{c}{\pi}\int_0^\infty \frac{\alpha_3(\Omega;\omega_2,\omega_3)}{\Omega^2 - \omega_1^2}d\Omega \qquad (11)$$

where $c$ is the light velocity in vacuum. In principle, this requires a nondegenerate $\alpha_3$. However, similar to the procedure given in Ref. [28] for the 2PA process, we approximate the nondegenerate $\alpha_3$ with the degenerate expression by using Eq. (10) and perform the KK transformation to obtain the wavelength dispersion of $n_4$, shown in Fig. 5. As an example, the values of $n_4$ for ZnO and ZnS at 780 nm are computed to be $-1.4 \times 10^{-26}$ cm$^4$/W$^2$ and $3.1 \times 10^{-27}$ cm$^4$/W$^2$, respectively. The contributions from the nonlinear processes, $n_2I$ and $n_4I^2$, are compared in Fig. 6.

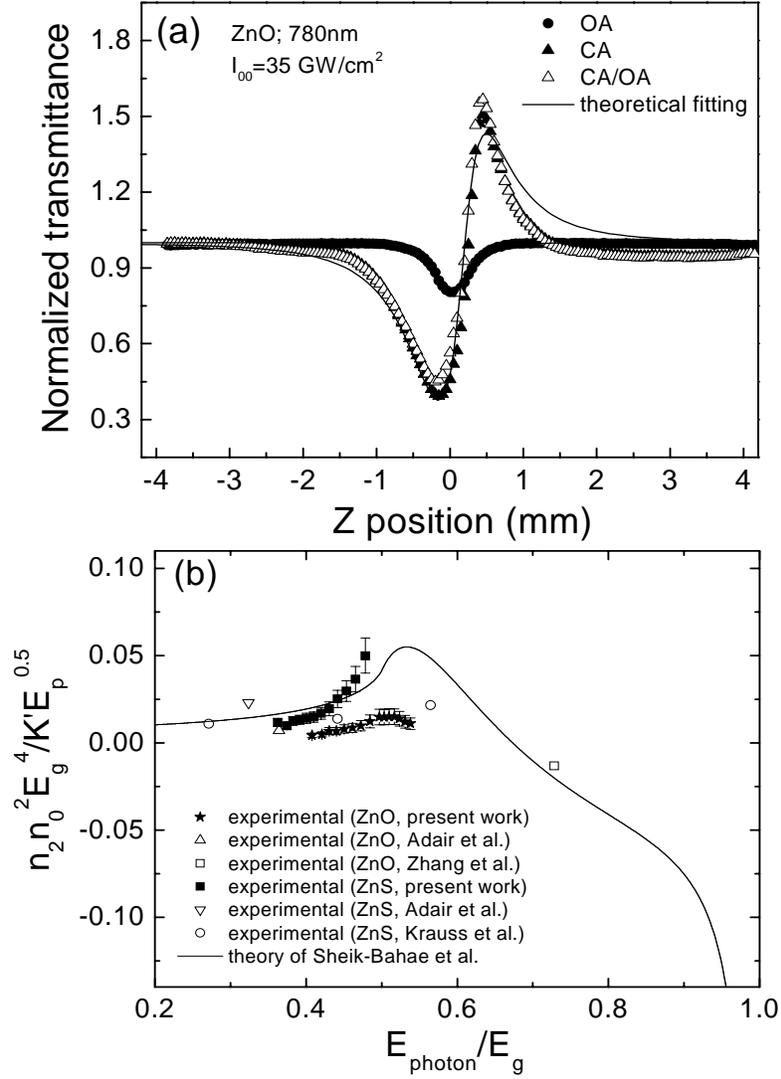

Fig. 4. (a) OA Z-scan, CA Z-scan and CA Z-scan divided by OA Z-scan; and (b) Kerr-type nonlinear refraction plotted as a function of $E_{photon}/E_g$. The solid scatters are the experimental data, while the solid lines are calculated from the theory of Sheik-Bahae *et al.* [26]. For comparison, the experimental results of Adair *et al.* (the empty triangles) [5], Zhang *et al.* (the empty square) [6], and Krauss *et al.* (the empty circles) [7], are also displayed.

We also calculate the nonlinear refractive effect induced by the 3PA-excited charge carriers, the number density of which is given by

$$\frac{dN_{e-h}}{dt} = \frac{\alpha_3 I^3}{3\hbar\omega} - \frac{N_{e-h}}{\tau} \tag{12}$$

where $\tau$ is the lifetime of the excited charge carriers, which is typically on the scale of 100 ps or longer [6]. The change in the refractive index due to $\sigma_r N_{e-h}$ is numerically calculated as a function of the laser irradiance, see Figs. 6(a) and 6(b). Within the irradiance range up to ~40 GW/cm$^2$, the ratio of $n_4I^2$ (or $\sigma_r N_{e-h}$) to $n_2I$ is estimated to be less than 10%, which indicates the fifth-order (or seventh-order) nonlinear refraction is insignificant under our experimental condition. Figures 6(c) and 6(d) reveal that the 3PA-excited free carrier absorption ($\sigma_a N_{e-h}$) is also negligible, compared to 3PA ($\alpha_3 I^2$) for both ZnO and ZnS under our experimental condition as well.

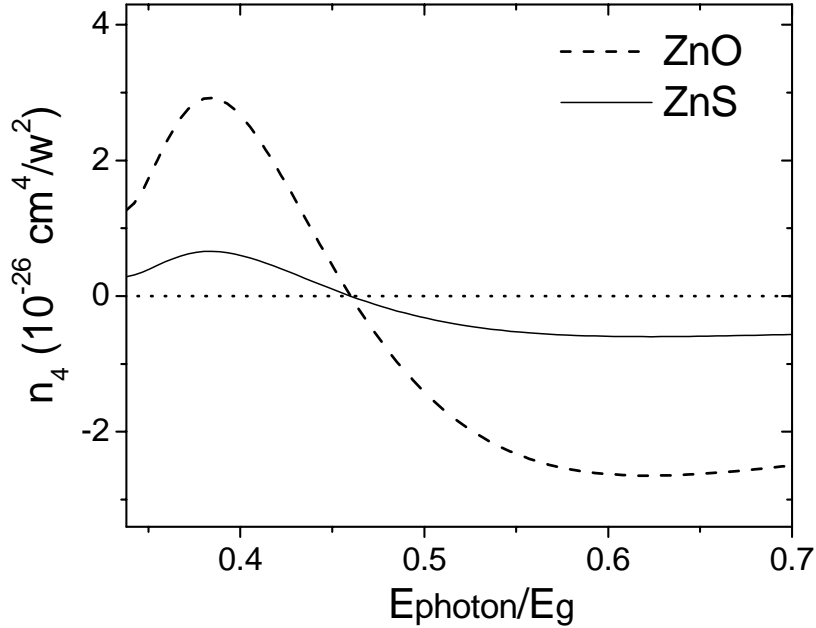

Fig. 5. Calculated nonlinear refraction $n_4$ plotted against $E_{photon}/E_g$, for both ZnO (the dashed line) and ZnS (the solid line).

*4.3 Transient transmission measurement*

Figure 7 shows the transient absorption measured with the pump-probe measurements on the ZnO and ZnS crystals. Again the free-carrier absorption induced by 3PA is found to be insignificant. Similarly, the femtosecond time-resolved OKE measurements lead to the same conclusion. It is also interesting to note that the autocorrelation signals in the pump-probe measurements are narrower than the autocorrelated OKE signal. This is understandable since the pump-probe signal is related to the 3PA, which is the fifth-order nonlinear mechanism, and the signal is proportional to $\int_{-\infty}^{t} I_{pump}^2(t') I_{probe}(t'-t) dt'$. But, the OKE signal originates from the Kerr-type refractive nonlinearity, which is the third-order nonlinear process and the signal is given by $\int_{-\infty}^{t} I_{pump}(t') I_{probe}(t'-t) dt'$. Figure 7 clearly illustrates that the 3PA dominates when the delay is between –200 fs and +200 fs. Beyond this region of near-zero delay, we see the signal is close to the autocorrelation due to 2PA. This should be expected since the crystals that we used are imperfect and there must be some defect states existing within the band gap. Two-photon transitions, for example between the defect states and the conduction band, are possible. The discrepancy in the slopes between experimental values obtained from the plotting of $Ln(1-T_{OA})$ vs. $Ln(I_0)$ in Fig. 1(b) and the theoretical value also indicate the co-existence of 2PA and 3PA processes. However, we estimate that the density of the defect states is insignificant, and the two-photon transitions should be saturated easily in the high excitation regime. Therefore, 2PA due to defect-state-related transitions should not play an important role in our Z-scans.

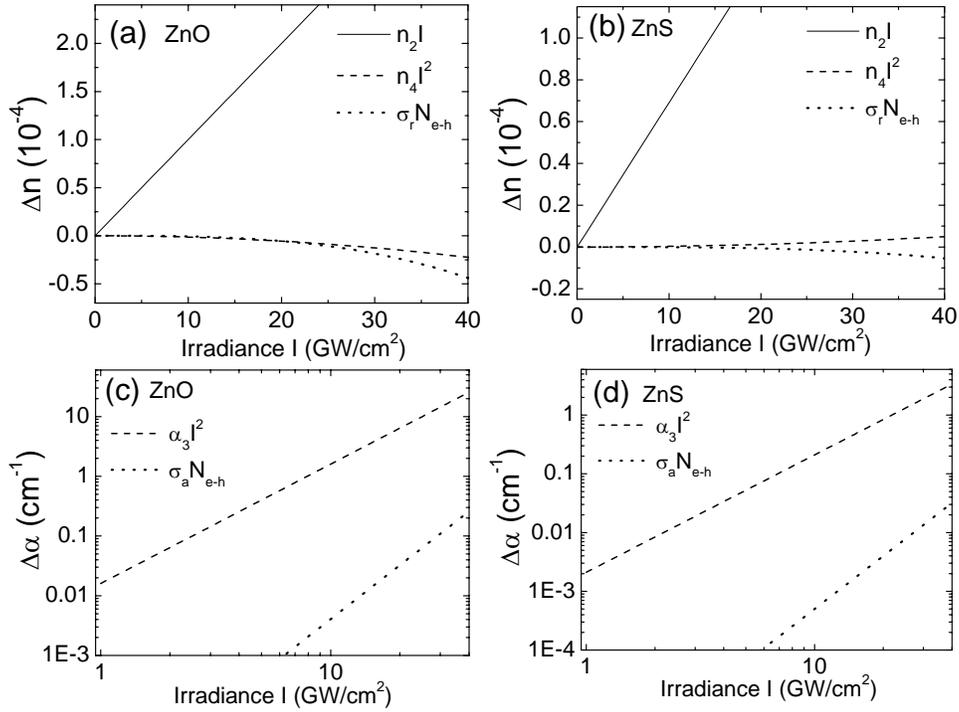

Fig. 6. Calculated nonlinear refraction changes ($n_2 I$, $n_4 I^2$, $\sigma_r N_{e-h}$) vs. the irradiance for (a) ZnO and (b) ZnS; and calculated changes in nonlinear absorption ($\alpha_3 I^2$, $\sigma_a N_{e-h}$) vs. the irradiance for (c) ZnO and (d) ZnS. Herein, wavelength is 780 nm, $n_2$ is $1.0 \times 10^{-5}$ cm$^2$/GW ($0.69 \times 10^{-5}$ cm$^2$/GW), $n_4$ is $-1.4 \times 10^{-26}$ cm$^4$/W$^2$ ($3.1 \times 10^{-27}$ cm$^4$/W$^2$), $\alpha_3$ is 0.016 cm$^3$/GW$^2$ (0.0021 cm$^3$/GW$^2$) for ZnO (ZnS) crystal, and $\sigma_r = -1.1 \times 10^{-21}$ cm$^3$ and $\sigma_a = 6.5 \times 10^{-18}$ cm$^2$ are assumed for both ZnO and ZnS crystals [6].

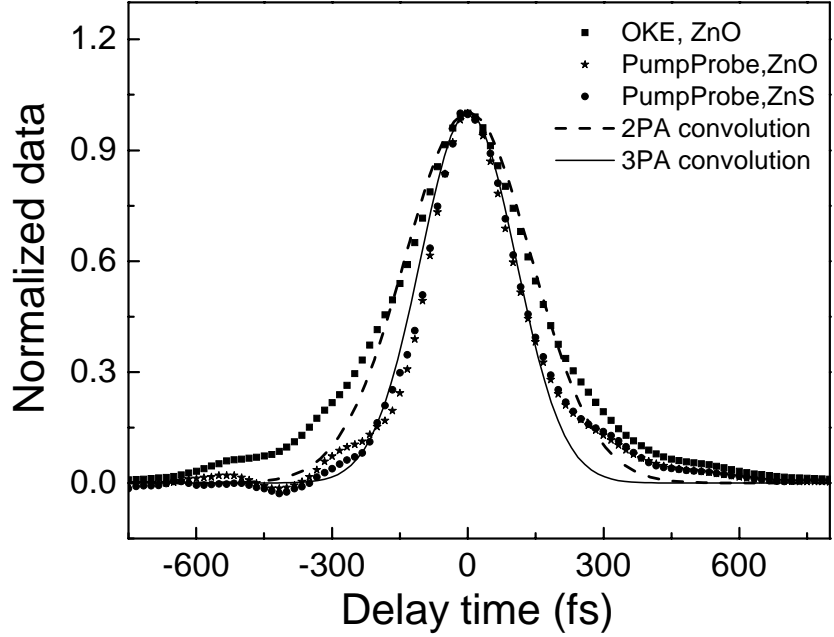

Fig. 7. Normalized OKE signal (the solid squares for ZnO) and pump-probe signals (the solid stars for ZnO and the solid circles for ZnS) measured at a wavelength of 780 nm as a function of the delay time. The experiment data were measured under the same excitation irradiance (~ 9.0 GW/cm$^2$) with 1-kHz pulse repetition rate. The dashed line and solid lines are the 2PA and 3PA intensity autocorrelation functions of the laser pulses, respectively.

*4.4 Figure of merit for optical switching*

It is important to be able to predict the influence of MPA on optical switching. In the absence of absorption, a phase change of $\pi$ in optical Kerr shutter switching can be obtained for a sufficient length of sample. Nonlinear absorption is not wanted as it may attenuate beam before the required nonlinear phase shift is achieved. In order to quantify the effects of 3PA, the nonlinear figure of merit (FOM) $V = I_c^2 \alpha_3 b_3 L_c$ has been formulated [14], where $I_c = \lambda/(b_2 n_2 L_c)$ is the cw critical intensity for switching and $L_c$ represents one half beat length. The symbols ($b_2$ and $b_3$) represent the overlap integrals over the model profiles in the waveguides for the third- and fifth-order nonlinear absorption, respectively. To calculate the nonlinear FOM of ZnO and ZnS, we assume the parameters as that in an AlGaAs nonlinear directional coupler: $L_c = 0.72$ cm, $b_2 = 1/2$ and $b_3 = 1/3$ [12]. All the results are listed in Table 1. At wavelengths of ~ 800 nm, the nonlinear FOM of ZnS is better than that of ZnO and meets the requirement of $V < 0.68$. It has been reported that $V$ is 0.1784 for AlGaAs waveguide at telecommunication wavelength of 1.55 $\mu$m [13]. For ZnO and ZnS, however, $V$ is zero theoretically at 1.3 $\mu$m or 1.55 $\mu$m, while $V$ is non-zero for ZnSe. And the two-band model, described in Ref. [26], predicts that the $n_2$ value (~1.0 × 10$^{-5}$ cm$^2$/GW) in ZnO is comparable with the one for ZnSe.

Table 1. Measured 3PA or 2PA coefficient, Kerr-type refractive nonlinearity, and calculated nonlinear FOM for ZnO and ZnS. The relative errors are estimated as ±20%

| λ (nm) | ZnS | | | ZnO | | |
|---|---|---|---|---|---|---|
| | $n_2$ ($10^{-5}$ cm$^2$/GW) | $\alpha_3$ cm$^3$/GW$^2$ | V | $n_2$ ($10^{-5}$ cm$^2$/GW) | $\alpha_3$ ($\alpha_2$) cm$^3$/GW$^2$ (cm/GW) | V |
| 720 | 1.4 | | | 0.75 | (1.3) | |
| 730 | | | | 0.83 | (1.0) | |
| 740 | 1.0 | | | 0.94 | (0.77) | |
| 750 | | | | 1.0 | (0.30) | |
| 760 | 0.81 | $1.9 \times 10^{-3}$ | 0.31 | 1.0 | (0.23) | |
| 770 | | | | 1.0 | (0.12) | |
| 780 | 0.69 | $2.1 \times 10^{-3}$ | 0.50 | 1.0 | 0.016 | 1.8 |
| 800 | 0.53 | $1.7 \times 10^{-3}$ | 0.70 | 0.84 | 0.010 | 1.8 |
| 820 | 0.46 | $1.2 \times 10^{-3}$ | 0.69 | 0.67 | $8.3 \times 10^{-3}$ | 2.3 |
| 840 | 0.40 | $0.70 \times 10^{-3}$ | 0.56 | 0.59 | $7.7 \times 10^{-3}$ | 2.9 |
| 860 | 0.39 | $0.60 \times 10^{-3}$ | 0.55 | 0.54 | $6.6 \times 10^{-3}$ | 3.1 |
| 880 | 0.36 | | | 0.46 | $4.1 \times 10^{-3}$ | 2.8 |
| 900 | 0.34 | | | 0.46 | $5.4 \times 10^{-3}$ | 3.8 |
| 920 | 0.27 | | | 0.33 | $4.6 \times 10^{-3}$ | 6.4 |
| 950 | 0.32 | | | 0.29 | $5.2 \times 10^{-3}$ | 10 |

## 5. Conclusion

In summary, we have measured the spectra of 3PA coefficients and Kerr-type nonlinear refraction from 720 nm to 950 nm in ZnO and ZnS crystals at laser excitation irradiances up to 40 GW/cm$^2$. The measured nonlinear refraction spectra are in good agreement with the two-band model, while the 3PA wavelength dependence is consistent with the theoretical prediction by Brandi and de Araujo as $(3E_{photon}/E_g-1)^{5/2}(3E_{photon}/E_g)^{-9}$. We have also evaluated higher-order nonlinear optical effects including the fifth-order nonlinear refraction associated with virtual three-photon transitions, and effectively seventh-order nonlinear processes induced by three-photon-excited free charge carriers. Furthermore, the nonlinear figure of merit $V$ is calculated and the ultrafast transient dynamics are investigated. Our results demonstrate that ZnS and ZnO have potential for optical switching applications.